# Lattice dynamics and negative thermal expansion in the framework compound ZnNi(CN)$_4$ with two-dimensional and three-dimensional local environments

Stella d'Ambrumenil,[1,2] Mohamed Zbiri,[1*] Ann M. Chippindale,[2] Simon J. Hibble,[3] Elena Marelli,[2] and Alex C. Hannon[4]

[1]Institut Laue-Langevin, 71 avenue des Martyrs, Grenoble Cedex 9, 38042, France
[2]Department of Chemistry, University of Reading, Whiteknights, Reading RG6 6AD, United Kingdom
[3]Chemistry Teaching Laboratory, Department of Chemistry, University of Oxford, South Parks Road, Oxford, OX1 3PS, United Kingdom
[4]ISIS Facility, Rutherford Appleton Laboratory, Chilton, Didcot, OX11 0QX, United Kingdom

[*]zbiri@ill.fr

## Abstract

ZnNi(CN)$_4$ is a three-dimensional (3D) framework material consisting of two interpenetrating PtS-type networks in which tetrahedral [ZnN$_4$] units are linked by square-planar [NiC$_4$] units. Both the parent compounds, cubic Zn(CN)$_2$ and layered Ni(CN)$_2$, are known to exhibit 3D and 2D negative thermal expansion (NTE), respectively. Temperature-dependent inelastic neutron scattering (INS) measurements were performed on a powdered sample of ZnNi(CN)$_4$ to probe phonon dynamics. The measurements were underpinned by *ab initio* lattice dynamical calculations. Good agreement was found between the measured and calculated generalized phonon density-of-states, validating our theoretical model and indicating that it is a good representation of the dynamics of the structural units. The calculated linear thermal expansion coefficients are $\alpha_a = -21.2 \times 10^{-6}$ K$^{-1}$ and $\alpha_c = +14.6 \times 10^{-6}$ K$^{-1}$, leading to an overall volume expansion coefficient, $\alpha_V$ of $-26.95 \times 10^{-6}$ K$^{-1}$, pointing towards pronounced NTE behavior. Analysis of the derived mode-Grüneisen parameters shows that the optic modes around 12 and 40 meV make a significant contribution to the NTE. These modes involve localized rotational motions of the [NiC$_4$] and/or [ZnN$_4$] rigid units, echoing what has previously been observed in Zn(CN)$_2$ and Ni(CN)$_2$. However, in ZnNi(CN)$_4$, modes below 10 meV have the most negative Grüneisen parameters. Analysis of their eigenvectors reveals that a large transverse motion of the Ni atom in the direction perpendicular to its square-planar environment induces a distortion of the units. This mode is a consequence of the Ni atom being constrained only in two dimensions within a 3D framework. Hence, although rigid-unit modes account for some of the NTE-driving phonons, the added degree of freedom compared with Zn(CN)$_2$ results in modes with twisting motions, capable of inducing greater NTE.

# I.    INTRODUCTION

Negative thermal expansion (NTE) is a counter-intuitive phenomenon attracting keen fundamental and applied scientific interest. NTE often occurs in framework materials in one, two or three dimensions [1]. With better understanding of its mechanism, there is the potential to tailor materials with a targeted thermal response or composite materials with an overall zero thermal expansion. Applications range from aerospace technology (for example, in telescope mirrors and optics exposed to extreme temperature in space) to biomedical engineering (for example, in prostheses, implants and dental fillings).

NTE in inorganic framework compounds stems from anharmonic lattice vibrations that shorten the mean distances between atoms [2]. Well-known NTE compounds include $Ag_2O$, $ReO_3$, $ScF_3$, $ZrW_2O_8$ and $Sc_2Mo_3O_{12}$ [1]. Due to the *tension effect*, neighboring metals are brought closer together by transverse displacements of the bridging ligands. An increase in temperature results in an increase in amplitude of these transverse motions, and hence a reduction in lattice parameter. For structures consisting of corner-sharing polyhedra, as in the compounds above, specific phonon modes exhibiting little or no polyhedral distortion give rise to the *tension effect*. These modes are known as rigid-unit-modes (RUMs), and are often related to NTE. As a diatomic ligand, such as C≡N, would reduce the structural constraints of a RUM, NTE is often observed in transition-metal cyanides [3, 4, 5, 6].

Metal cyanides exhibiting NTE include $Zn(CN)_2$ and $Ni(CN)_2$ [4, 5]. $Zn(CN)_2$ adopts a cubic structure consisting of two interpenetrating diamondoid frameworks formed from corner-sharing tetrahedra [5]. It exhibits three-dimensional (3D) NTE with a linear thermal expansion coefficient, $\alpha_a$, of $-16.9 \times 10^{-6}$ $K^{-1}$, corresponding to a volume thermal expansion, $\alpha_V$, of $-51 \times 10^{-6}$ $K^{-1}$ [7]. In contrast, $Ni(CN)_2$ has a layered structure consisting of $Ni^{2+}$-centred corner-sharing square-planar units. This leads to 2D NTE in the plane of the layers with $\alpha_a$ equal to $-6.5 \times 10^{-6}$ $K^{-1}$, where $a = b$ is the in-plane lattice parameter, and a positive thermal expansion perpendicular to the layers resulting in an overall positive volume expansion, $\alpha_V$, of $+ 48 \times 10^{-6}$ $K^{-1}$ [4]. In both compounds, the dimensionality of the NTE follows the dimensionality of the structure, a feature also displayed by the Group 11 metal cyanides (e.g. CuCN, AgCN and AuCN), which have structures based on 1D chains and show 1D NTE [3, 8]. Inelastic neutron scattering (INS) and *ab initio* studies of $Ni(CN)_2$ and $Zn(CN)_2$ have confirmed the existence of low-energy phonon modes with negative Grüneisen parameters, which give rise to NTE [9]. In $Zn(CN)_2$, Fang *et al.* have shown that there is a strong correlation between a mode having a negative Grüneisen parameter and its resemblance to a pure RUM [10].

The structures of $Ni(CN)_2$ and $Zn(CN)_2$ contain metals in only one type of coordination geometry, with $Ni^{2+}$ in a 2D (square-planar) and $Zn^{2+}$ in a 3D (tetrahedral) environment. Recent studies combining 3D units (tetrahedral or octahedral) with pseudo 1D units (from Group 11 metals) within the same compound have resulted in materials with three-dimensional structures which have very large linear thermal expansion coefficients. Examples include $ZnAu_2(CN)_4$ and $Ag_3[Co(CN)_6]$ [11, 12, 13]. The lower constraint of a 1D metal environment within a 3D framework naturally induces phonons with large RUM-like motions.

This work reports on the lattice dynamics of $ZnNi(CN)_4$, which contains metals in both 2D and 3D local environments [14]. The tetragonal structure, as reported by Yuan *et al.* [14], consists of two interpenetrating cooperite (PtS) type $ZnNi(CN)_4$ frameworks (Fig. 1). A second crystal structure, also tetragonal, but consisting of a single $ZnNi(CN)_4$ framework was later reported [15].

The dual metal-environment dimensionality of $ZnNi(CN)_4$ provides a different and interesting system for studying phonon dynamics and its relationship to NTE. Indeed, another PtS-type framework compound, $Cd(NH_3)_2[Cd(CN)_4]$, has recently been discovered as an NTE material [16]. Here we present a detailed lattice dynamical study of $ZnNi(CN)_4$ using INS and *ab initio* calculations. In addition, total neutron diffraction, powder x-ray diffraction (XRD), infra-red (IR) and Raman spectroscopy data are also reported and discussed.

## II.    EXPERIMENTAL DETAILS

A 2-g sample of $ZnNi(CN)_4$ was synthesized as follows: 50-ml aqueous solution of $K_2Ni(CN)_4 \cdot n H_2O$ (Aldrich) (5.2808 g, 0.022 mol) was added to 50 ml of $Zn(NO_3)_2 \cdot 6H_2O$ (BDH) (6.5199 g, 0.022 mol). The resulting gelatinous precipitate, cream in color, was stirred for 3 h, filtered, and repeatedly washed with distilled water. The wet product was then dried in a vacuum oven for 1 h at 373 K.  The sample density was measured at room temperature using a Quantachrome micropycnometer with helium gas as the working fluid. IR frequencies were measured with a Perkin Elmer Spectrum 100 spectrometer and Raman frequencies with a Renishaw InVia Raman microscope ($\lambda_{exc} = 785$ nm). A powder x-ray diffractogram was collected in transmission mode at room temperature using a Bruker D8 diffractometer equipped with primary monochromatic Cu $K_{\alpha 1}$ radiation ($\lambda = 1.54060$ Å). The sample was finely ground and loaded into a quartz capillary (0.9 mm O.D.)

Total neutron diffraction data were collected at 15 and 295 K using the time-of-flight GEM diffractometer (ISIS Neutron and Muon Source, Didcot), designed to collect data over a wide-angle range ($1.2 \leq 2\theta \leq 171.4°$). The interference function, $Qi(Q)$, was obtained by merging data from detector banks $2 - 5$ with a $Q_{max}$ of 40 Å$^{-1}$ and extrapolating the low-$Q$ region with a function of the form $a + bQ^2$ (where $a$ and $b$ are constants). The total neutron correlation function, $T_N(r)$, was obtained by Fourier transform of $Q_i(Q)$, using the Lorch function [17, 18] to reduce the termination ripples. The bond lengths were extracted by fitting the first peaks in the $T_N(r)$ shown in Fig. 2 using the method described elsewhere [19, 20].

INS measurements were performed on 2-g sample of $ZnNi(CN)_4$, using the cold-neutron, time-of-flight, time-focusing, IN6 spectrometer (Institut Laue-Langevin, Grenoble), operating in the high-resolution mode, and offering a good signal-to-noise ratio. The IN6 spectrometer supplies a typical flux of $10^6$ n cm$^{-2}$ s$^{-1}$ on the sample, with a beam-size cross section of 3×5 cm$^2$ at the sample position. The sample was placed inside a thin-walled aluminium container and fixed to the tip of the sample stick of an orange cryofurnace. An optimized small sample thickness of 1 mm was used, to minimize effects such as multiple scattering and absorption. An incident wavelength of 4.14 Å was used, with an elastic energy resolution of 0.17 meV, as determined from a standard vanadium sample. The vanadium sample was also used to calibrate the detectors. Data were collected up to 100 meV in the up-scattering, neutron energy-gain mode, at 200, 300 and 400 K.

On IN6, under these conditions, the resolution function broadens with increasing neutron energy, and it can therefore be expressed as a percentage of the energy transfer. The ILL program LAMP [21] was used to carry out data reduction and treatment, by performing detector efficiency calibration, background subtraction and multiphonon correction [22]. Background reduction included measuring an identical empty container in the same conditions as sample measurements. At the used shortest-available neutron wavelength on IN6, $\lambda$ = 4.14 Å, the IN6 angular coverage ($\sim 10 - 114°$) corresponds to a maximum momentum transfer of $Q \sim 2.6$ Å$^{-1}$.

The $Q$-averaged, one-phonon generalized phonon density of states (GDOS) [23], $g^{(n)}(E)$, was obtained using the incoherent approximation [25, 26]. This method has been used previously to investigate phonon dynamics and anomalous thermal expansion in a number of transition-metal cyanides [9, 11, 13, 47, 27]. In the incoherent, one-phonon approximation, $g^{(n)}(E)$ is related to the measured scattering function $S(Q, E)$ from INS by

$$g^{(n)}(E) = A \left\langle \frac{e^{2W(Q)}}{Q^2} \frac{E}{n(E,T) + \frac{1}{2} \pm \frac{1}{2}} S(Q, E) \right\rangle \qquad (1)$$

where $A$ is a normalization constant, $2W(Q)$ is the Debye-Waller factor and $n(E,T)$ is the thermal occupation factor equal to $[\exp(E/k_BT{-}1)]^{-1}$. The + or − signs correspond to neutron energy loss or gain respectively and the bra-kets indicate an average over the whole $Q$ range [28].

## III.   COMPUTATIONAL DETAILS

Projector-augmented wave potentials [29, 30] calculations were carried out using the Vienna *ab initio* simulation package (VASP) [31 − 34]. The generalized gradient approximation was adopted within the Perdew-Burke-Ernzehof scheme [35, 36]. An energy cut-off of 700 eV was set and Gaussian broadening implemented with a smearing width of 0.01 eV.

The tetragonal crystal structure (Fig. 1) published by Yuan *et al.* [14] was used as an input structural model for the *ab initio* calculations. Total structural relaxation, followed by constant volume relaxation at seven different volumes close to equilibrium, was performed using a 5 × 5 × 2 *k*-point mesh generated using the Monkhorst-Pack method [37].

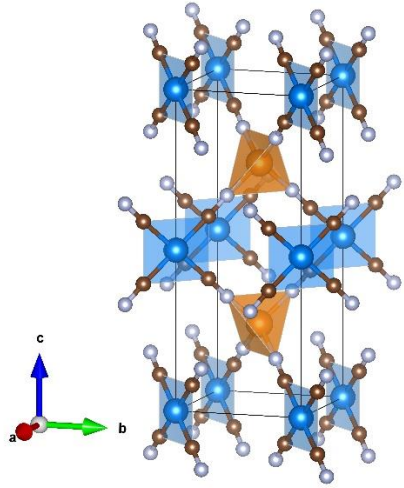

*FIG. 1. Structural model of ZnNi(CN)₄ (showing the square-planar Ni(CN)₄ and tetrahedral Zn(NC)₄ units), which was used for ab initio calculations [14]. Key: Ni, blue, Zn, orange, C, brown, N, light blue.*

The results of total energy calculations at varying volumes were fitted to a Birch-Murnaghan equation of state to establish the bulk modulus, $B_0$, and its pressure derivative, $B_0$', defined as,

$$B_0 = -\left(\frac{\partial \ln V}{\partial P}\right)^{-1}_{T,P=0} \qquad B_0' = -\left(\frac{\partial B}{\partial P}\right)^{-1}_{T,P=0} \qquad (2)$$

For comparison, bulk modulus calculations were also carried out for the parent compounds, $Zn(CN)_2$ and $Ni(CN)_2$, using the crystal structures published in [5] and [4], respectively. Due to the layered nature of $Ni(CN)_2$, the van der Waals-based density functional scheme, vdW-DF2 [38 − 41], was used to account for possible weak interactions between neighboring layers.

Phonon calculations using the direct method [42] were performed for each volume on a $2 \times 2 \times 2$ supercell with a $2 \times 2 \times 1$ $k$-point mesh. The calculated partial vibrational density-of-states (PDOS) of the $k^{\text{th}}$ atom contributes to the measurable GDOS via,

$$g^{(n)}(E) = B \sum_k \left(\frac{4\pi b_k^2}{m_k}\right) g_k(E) \qquad (3)$$

where $B$ is a normalization constant, $b_k$ is the neutron scattering length and $m_k$ is the mass of the $k^{\text{th}}$ atom [28]. The constant $(4\pi b_k^2/m_k)$ represents the neutron weighting factor of the $k^{\text{th}}$ atom with values: Ni, 0.3152, Zn, 0.0632, C, 0.4622, N, 0.8216 barns amu⁻¹ [43]. The mean square displacement (MSD) of the $k^{\text{th}}$ atom in the $\hat{x}$ direction at temperature, $T$, is related to the PDOS projected along $\hat{x}$ as follows:

$$\langle |u_k^x|^2 \rangle = \int \frac{\hbar}{m_k \omega} \left(\frac{1}{2} + n(\omega, T)\right) g_k^x(\omega) d\omega \qquad (4)$$

The integrand in equation (4) represents the contribution of modes with energy $\hbar\omega$ to the MSD.

Thermal expansion stems for anharmonic phonon modes, whose frequencies changes with temperature [2]. The expansion is negative when the sum over all mode Grüneisen parameters, defined as $\gamma_i = -(d\ln\omega_i/d\ln V)$, is also negative [44]. The total anharmonicity can be split into two terms as follows,

$$\left(\frac{\partial \omega_i}{\partial T}\right)_P = \left(\frac{\partial \omega_i}{\partial T}\right)_V + \left(\frac{\partial \omega_i}{\partial V}\right)_T \left(\frac{\partial \omega_i}{\partial T}\right)_P \qquad (5)$$

$$= \left(\frac{\partial \omega_i}{\partial T}\right)_V - \alpha \omega_i \gamma_i^T \qquad (6)$$

where $\omega_i$ is the mode frequency, $\alpha$ is the volume expansion coefficient and $\gamma_i^T$ is the isothermal Grüneisen parameter [45]. The first term, where the frequency is changing with temperature at constant volume, is the explicit anharmonicity sometimes referred to as the true anharmonicity. The second term is the implicit anharmonicity and is due to volume effects.

The quasiharmonic approximation (QHA), where phonon calculations at different volumes are used to mimic a temperature effect [2], was used to calculate the thermal expansion coefficient and mode Grüneisen parameters. Equation 5 and 6 can be rearranged as follows,

$$\gamma_i^P = \gamma_i^T - \frac{1}{\alpha \omega_i}\left(\frac{\partial \omega_i}{\partial T}\right)_V \qquad (7)$$

where $\gamma_i^P$ is the isobaric Grüneisen parameter. In the QHA, $\gamma_i^T$ is equal to $\gamma_i^P$ as the explicit anharmonic effects are ignored [45]. The QHA is valid for low temperatures [46] and has frequently been used to compute accurate negative thermal expansion coefficients using calculated Grüneisen parameters [9, 11, 13, 47 − 53]. However, there are limitations to the method when explicit anharmonicity is dominant [54, 55]. It is worth noting that throughout this paper, calculated Grüneisen parameters are, in fact, isothermal Grüneisen parameters. The phonon density-of-states, mode frequencies and Grüneisen parameters were determined using the Phonopy software [56].

## IV.   RESULTS AND DISCUSSION
### A. Structure

A single crystal of the doubly interpenetrating $ZnNi(CN)_4$ structure was synthesized by Yuan *et al.* via slow diffusion of $Zn(SO_4)$ and $K_2[Ni(CN)_4]$ in a mixture of ethanol and water [14]. A second crystal structure consisting of a single $ZnNi(CN)_4$ framework was later synthesized by a similar method [15]. It was noted for both structures that, when removed from the mother liquor, the crystals disintegrated to a powder [14, 15]. In the case of the single-network structure, this is likely to be due to loss of occluded solvent molecules. It is unclear as to why crystals of the double framework should pulverize upon removal from the solvent, but it correlates with our observation that powder samples of the doubly interpenetrating form are poorly crystalline, as shown by the XRD pattern (Fig. 2). INS measurements were of necessity carried out on a large, solvent-free, and therefore poorly crystalline sample of $ZnNi(CN)_4$.

The sample density measured at room temperature was 2.0539 g cm$^{-3}$, comparable to that obtained for the double interpenetrating framework (2.106 g cm$^{-3}$) [14], and almost double the density calculated for the single-framework material (1.047 g cm$^{-3}$) [15]. The peak positions in the powder XRD pattern match those of the doubly-interpenetrating structure. However, the broad peaks renders measurement of thermal expansion coefficients impossible by refinement of unit-cell parameters at varying temperature.

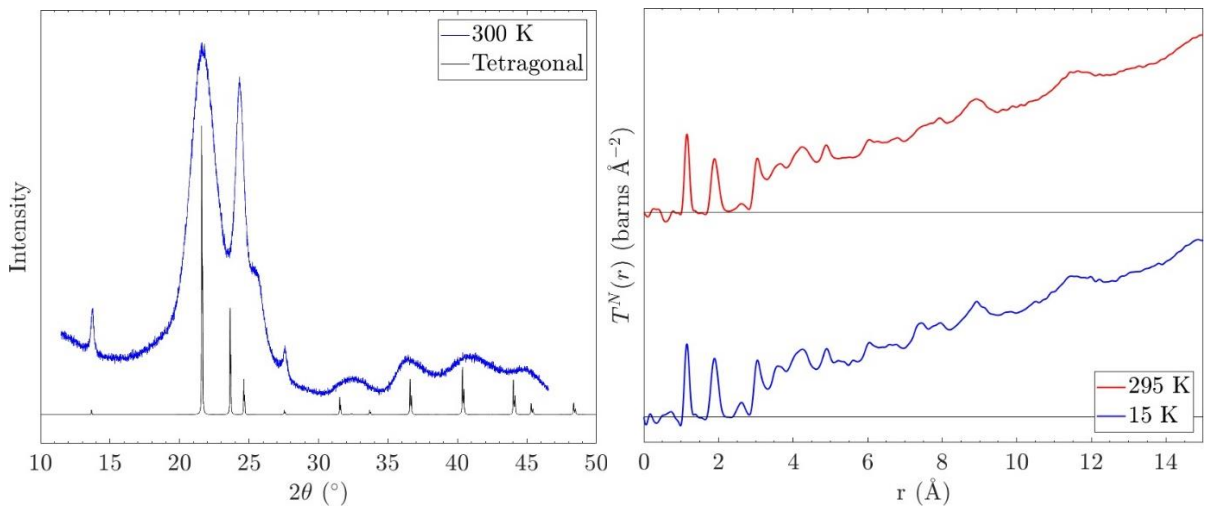

FIG. 2. *Left: Comparison of the measured x-ray powder diffraction pattern (blue) with a theoretical pattern calculated from the tetragonal double interpenetrating structure [14]. Right: The total correlation function of ZnNi(CN)$_4$ at 15 and 295 K obtained from Fourier transformation of the $Q_i(Q)$ measured by total neutron diffraction.*

Structural relaxation of the room-temperature crystal structure obtained by Yuan *et al.* [14] converged easily. In the relaxed structure, there was a slight deviation from linearity of the Zn-NC-Ni linkage, with the sum of the individual bond lengths being 0.02 Å larger than the Zn…Ni distance. Such a deviation reduces the decreasing effect any transverse motion of the ligand would have on the unit-cell parameters. Nevertheless, the unit-cell volume after the density functional theory (DFT)-based structural relaxation (at 0 K) increased by 6 Å$^3$, with an increase in the *a* parameter and slight decrease in the *c* parameter (Table I).

The single-crystal x-ray diffraction study by Yuan *et al.* [14] reported that the cyanide ligand bonds to Ni exclusively through C rather than N [9]. This orientation was confirmed by analysis of the measured pair distribution functional (PDF). Total energy calculations of the relaxed and unrelaxed crystal structure were carried out and then repeated with the CN ligand flipped. In both cases, an energy difference of ~3 eV between the two configurations was found, providing further confirmation of the original atom assignment and the formation of Ni–C≡N–Zn linkages.

The bond lengths resulting from analysis of the neutron correlation function, $T_N(r)$, are listed in Table I, together with those from [14] and those resulting from our structural relaxation. There is a significant discrepancy between the bond-length values from [14] and those extracted from the $T_N(r)$, the former being unusual and possibly incorrect. The bond lengths derived from our theoretical model are much closer to the values from the total neutron diffraction study and correctly identify the Ni-C bond as the shorter of the two single bonds.

Table I. *Bond lengths extracted from the analysis of our neutron pair distribution function measurements (n-PDF), shown in Fig. 2, together with those from the crystal structure of Yuan et al. [14] and those from the relaxed computational model. For the latter two the unit-cell parameters are also listed. The relaxed structural model correctly identifies that the Zn-N bond is longer than the Ni-C bond.*

| Bond Length / Å | $n$-PDF 295 K | $n$-PDF 15 K | Yuan Structure 300 K [14] | Relaxed Structure 0 K |
|---|---|---|---|---|
| C≡N | 1.159(1) | 1.153(1) | 1.128 | 1.169 |
| Ni-C | 1.870(5) | 1.857(3) | 1.921 | 1.835 |
| Zn-N | 1.980(6) | 1.970(4) | 1.892 | 1.968 |
| Unit-cell parameter $a$ /Å | | | 5.2629(15) | 5.32 |
| Unit-cell parameter $c$/Å | | | 12.987(8) | 12.95 |

The bulk modulus, $B_0$ and its pressure derivative, $B_0'$, were calculated for $ZnNi(CN)_4$ and its parent compounds, $Zn(CN)_2$ and $Ni(CN)_2$. The results together with other values from the literature are listed in Table II.

Table II. *Values of the bulk modulus, $B_0$, and its pressure derivative, $B_0'$, for $ZnNi(CN)_4$ and its parent compounds, $Ni(CN)_2$ and $Zn(CN)_2$. A comparison is given of DFT calculations carried out in this study with those from previous studies and experimentally acquired data at 300 K. The larger values for $B_0$ and $B_0'$ from the DFT calculations compared to those experimentally derived is expected, due to the temperature-dependence of these properties [57].*

| | This work | | | Previous Calculations | | Experimental 300 K | |
|---|---|---|---|---|---|---|---|
| | $Ni(CN)_2$ | $Zn(CN)_2$ | $ZnNi(CN)_4$ | $Ni(CN)_2$ | $Zn(CN)_2$ | $Ni(CN)_2$ | $Zn(CN)_2$ |
| Bulk Modulus, $B_0$ / GPa | 18.4 | 44.9 | 41.8 | 63.4 [9] | 84.1 [9], 59 [45], 46.4 [57], 45.0 [58] | 12.4 [59] | 34.2 [60], 33.4 [57], 25±11 [61] |
| Bulk Modulus Derivative, $B_0'$ | 9.2 | 4.6 | −2.5 | | 7.2 [57] | 4 [59] | −4.2 [57] |

A study by Fang *et al.* of the temperature dependence of the bulk properties of $Zn(CN)_2$ reported measurements of $B_0$ and $B_0'$ between 30 and 300 K [57]. These revealed that $B_0$ increased with decreasing temperature by 5.77 GPa over the investigated temperature range. However $B_0'$ decreased from −4 at 300 K to −9 at 250 K and then increased slowly, reaching −2 at 30 K. This behavior was then reproduced by the authors using molecular-dynamics simulations. The temperature dependence accounts for the difference between the values from our DFT calculations (0 K) and the experimentally measured values (300 K) in Table II.

Our values of $B_0$ and $B_0'$ for Ni(CN)$_2$ are also slightly higher than the experimental values, and this is also likely to arise from temperature dependence. Our value for $B_0$ is much closer to the experimental value than that from a previous study in which calculations were performed only on a single layer of Ni(CN)$_2$ which did not take into account inter-layer interactions [9].

The calculated value of $B_0$ for ZnNi(CN)$_4$ is much closer to that of Zn(CN)$_2$ than that of Ni(CN)$_2$. This is not unexpected as the structures of ZnNi(CN)$_4$ and Zn(CN)$_2$ both contain two interpenetrating 3D frameworks, in contrast to the layered structure of Ni(CN)$_2$. The calculated value of $B_0'$ for ZnNi(CN)$_4$ is lower than that of both parent compounds. Furthermore, the value is negative meaning that, even close to 0 K, the compound undergoes anomalous softening of the elastic modulus.

## B. Dynamics

The measured GDOS [23] of ZnNi(CN)$_4$, at 200, 300 and 400 K, are shown in Fig. 3. The spectra contain several well-pronounced vibration bands, whose energies do not exhibit any detectable temperature dependence in terms of frequency shifts. A low-energy shoulder observed around 6.5 meV decreases in intensity with temperature. It is located between the lowest-energy bands seen in the phonon spectra of Zn(CN)$_2$ and Ni(CN)$_2$, at 4 meV and 10 meV, respectively [9]. A large band between 7.5 and 20.0 meV contains several small features. The instrumental resolution does not allow their positions to be accurately determined. Other prominent bands are at 24, 41, 58 and 68 meV, along with two smaller bands at 32 and 51 meV. The latter are most clearly observed at 200 K due to the reduced Debye-Waller broadening at lower temperature. This justifies the use of the spectrum acquired at 200K to compare hereafter with the calculation ZnNi(CN)$_4$. Fig. 4 shows the calculated GDOS for ZnNi(CN)$_4$ which reproduced the measured GDOS well, in terms of both peak positions and intensities. This good agreement provides a further validation of the use of the ordered tetragonal model for lattice dynamical studies.

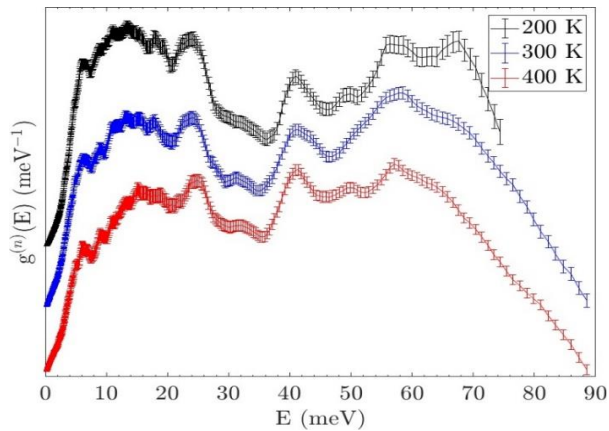

FIG. 3. *The generalized phonon density-of-states (GDOS) [23] of ZnNi(CN)$_4$ at 200, 300 and 400 K. Measurements were carried out with an incident wavelength of 4.14 Å on the IN6 spectrometer at the ILL. The spectra are vertically shifted with respect to each other to aid comparison.*

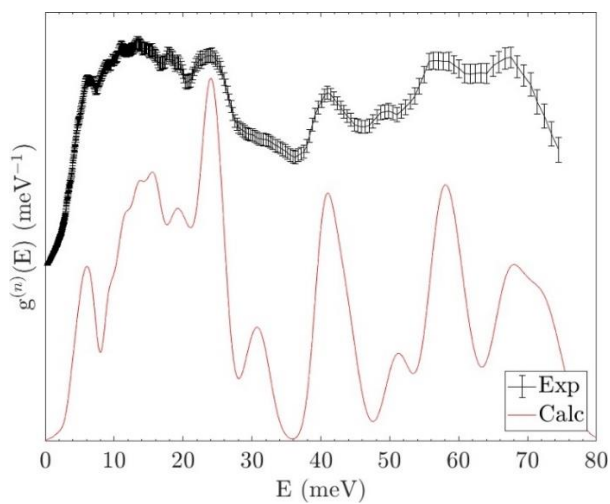

FIG. 4. *Comparison between the experimental (Exp), at 200 K, and calculated (Calc) phonon spectra of ZnNi(CN)₄. The calculated spectrum has been convoluted with a Gaussian of a full width at half maximum of 8% of the energy transfer to describe the effect of the experimental energy resolution.*

The calculated, neutron-weighted, partial vibrational density-of-states (PDOS) of each atomic species projected along the $x$ ($y$) and $z$ directions contributing to the total, are shown in Fig. 5. The calculated C≡N stretching modes lie between 270 and 280 meV ($2170 - 2260$ cm$^{-1}$), which are in the same region as the stretching frequencies found in the IR for $Zn(CN)_2$ (2218 cm$^{-1}$) and $Ni(CN)_2$ (2202, 2215 cm$^{-1}$) [9]. The figure shows that dynamics involving Ni atoms moving primarily in the $xy$ plane contribute the most to the low-energy feature seen in the GDOS at around 7 meV. In addition, it reveals that Zn moves very little even compared to Ni, indicating it is well anchored within the structure.

Below 40 meV, N contributes more to the total density-of-states than C. This observation can be interpreted as a greater freedom of movement around the tetrahedrally-coordinated Zn. Above 40 meV, the peaks are sharp and well defined indicating that these vibrations are localized and hence will resemble the vibrations of isolated tetrahedral and square-planar units. As the symmetry of the unit cell is the same as that of the [NiC₄] unit, all the motions of a structural unit with D$_{4h}$ symmetry are possible (A$_{2u}$, B$_{1g}$, B$_{2g}$ and E$_u$), however only the E vibration of the tetrahedral unit will be present, as the T$_2$ vibration will be lost due to the lack of three-fold symmetry. This would explain the larger contribution of C than N in this region. The peaks at 68 and 72 meV are from stretches of the Zn-N and Ni-C single bonds. The higher-energy of the two involves strong movement of the Ni and hence corresponds to the asymmetric stretch characterized by the irreducible representation, E$_u$, within the D$_{4h}$ local symmetry. The peaks between 50 and 65 meV are likely to arise from localized deformations of the units. The similarity of the C and N peak profiles between 40 and 50 meV, along with the almost stationary state of the metals, indicate a symmetry between the motions of the different units, which could correspond to rotational RUMs.

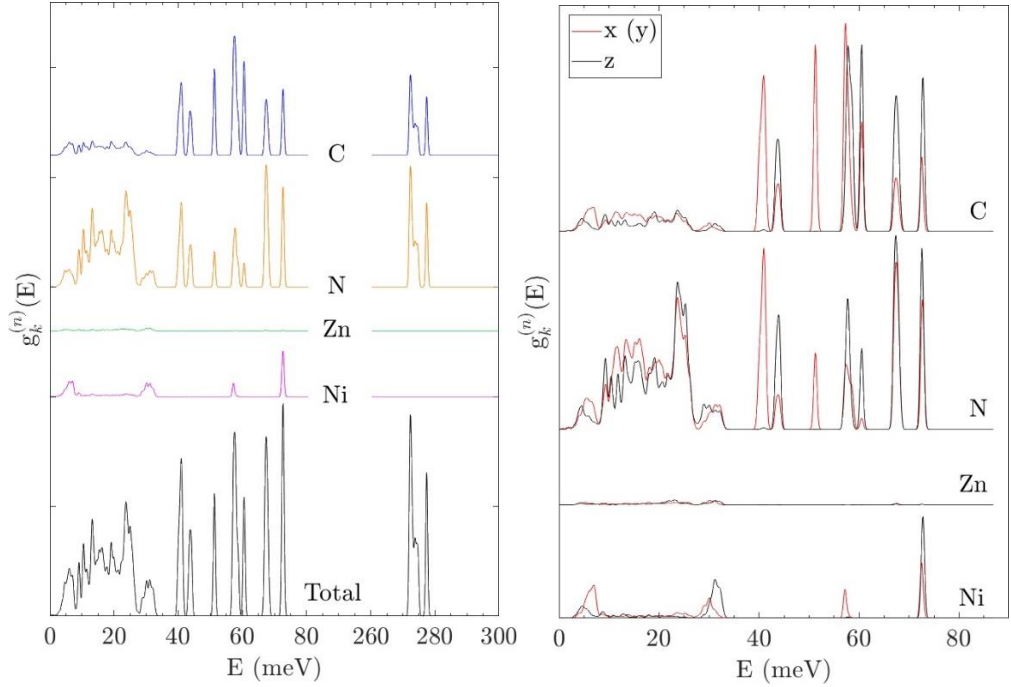

FIG. 5. *Left: The calculated PDOS for each atom in ZnNi(CN)₄. N contributes the most below 30 meV, however C and Ni are dominant in the lowest-energy phonon band at 7 meV. Right: The calculated PDOS for each atom projected onto the x (y) and z axes. The CN stretching modes are omitted for clarity in the right panel.*

The measured and calculated IR and Raman frequencies are listed in Table III. There are 42 optic modes, of which 15 are doubly degenerate. The structure belongs to the $4/mmm$ Laue class ($D_{4h}$ point group) and therefore the principle of mutual exclusion applies. This results in 13 IR and 13 Raman frequencies along with 16 silent modes; $\Gamma_{optic} = 4A_{1g} + 2\ A_{1u} + 2A_{2g} + 5A_{2u} + 2B_{1g} + 5B_{1u} + 5B_{2g} + 2B_{2u} + 7E_g + 8E_u$.

Table III. *Measured and calculated vibrational frequencies (cm⁻¹) from IR and Raman data.*

|  | IR active (13) | | Raman active (13) | | | Silent (16) | | | | |
|---|---|---|---|---|---|---|---|---|---|---|
| $\Gamma$ | $5A_{2u}$ | $8E_u$ | $4A_{1g}$ | $2B_{1g}$ | $7E_g$ | $2A_{1u}$ | $2A_{2g}$ | $5B_{1u}$ | $5B_{2g}$ | $2B_{2u}$ |
| Calculated | 197 | 56 | 202 | 154 | 73 | 186 | 98 | 145 | 176 | 125 |
|  | 266 | 138 | 490 | 335 | 109 | 416 | 321 | 241 | 221 | 412 |
|  | 458 | 199 | 548 |  | 178 |  |  | 472 | 490 |  |
|  | 588 | 240 | 2239 |  | 327 |  |  | 590 | 547 |  |
|  | 2197 | 460 |  |  | 349 |  |  | 2214 | 2237 |  |
|  |  | 465 |  |  | 541 |  |  |  |  |  |
|  |  | 584 |  |  | 2206 |  |  |  |  |  |
|  |  | 2194 |  |  |  |  |  |  |  |  |
| Measured | 573, 2190 | | 152, 172, 207, 324, 2197, 2214 | | | | | | | |

Having established that phonon calculations at equilibrium volume reproduced our measurements well, further calculations at varying volume were undertaken to extract mode Grüneisen parameters along with thermal properties of the ordered structure under the QHA. The calculated linear thermal expansion coefficients are $\alpha_a = -21.2 \times 10^{-6}$ K$^{-1}$ and $\alpha_c = +14.6 \times 10^{-6}$ K$^{-1}$, leading to an overall volume expansion coefficient of $\alpha_V = -26.95 \times 10^{-6}$ K$^{-1}$ between 200 and 400 K. This value of $\alpha_V$ is approximately half that measured and calculated for Zn(CN)$_2$ and in contrast to the overall positive volume expansion observed for Ni(CN)$_2$. However, the calculated value of $\alpha_a$ is more negative than the values determined for both Zn(CN)$_2$ and Ni(CN)$_2$ [9].

The variation of the volume thermal expansion coefficient with temperature and the contribution of phonon modes of energy E to the expansion are presented in Figs. 6 and 7, respectively. The expansion coefficient decreases with temperature from 0 K, reaching its most negative value slightly below 200 K. This value remains constant up to 400 K, above which it increases slowly. The change occurring in the first 200 K is the result of increasing contribution from phonon modes at 40 meV with temperature. This behavior is similar to that calculated for Zn(CN)$_2$ where optic modes at 40 meV, which are pure rotational RUMs, provide a large contribution to the NTE [10].

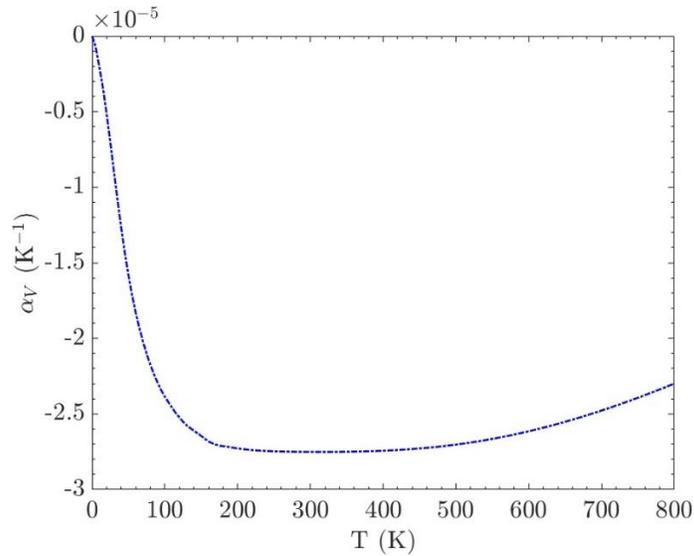

FIG. 6. *Volume thermal expansion coefficient, $\alpha_V$, vs temperature in ZnNi(CN)$_4$, calculated using the QHA.*

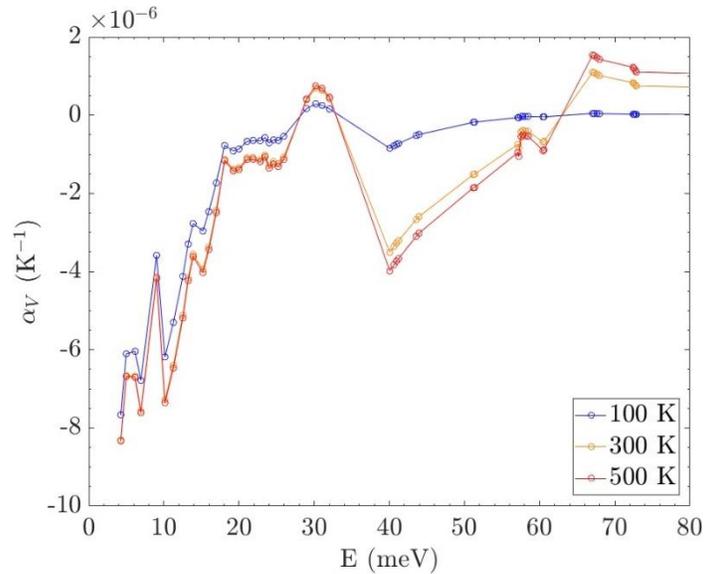

FIG. 7. *Contribution of modes of energy E to the thermal expansion in ZnNi(CN)₄.*

The calculated thermal ellipsoids at 300 K are shown in Fig. 8. The Ni and C atoms move predominantly in the direction perpendicular to their square-planar environment, whereas the N atoms move isotropically around the bond and Zn moves very little. The MSD of each atom with temperature is shown in Fig. 9. The figure also shows the contribution to the MSD with energy for each atom at 300 K, as well as the contribution projected in the $x$ ($y$) and $z$ directions averaged over all atoms.. It is clear that N has a higher amplitude of motion than C, as a result of contributions from modes between 10 and 20 meV. Similarly, the amplitude of motion of Ni is larger compared to Zn. At 7 meV, Ni even has a larger MSD than both light atoms. The motion of Ni at this energy will be in the $xy$ plane, perpendicular to the square-planar geometry as seen in Fig. 8.

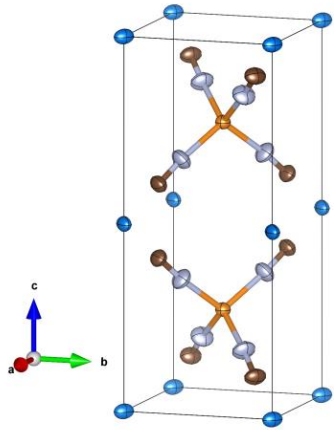

FIG. 8. *Relaxed structure of ZnNi(CN)₄ with calculated thermal ellipsoids at 300K representing a 75% probability of containing the atom. The Ni atom has a large oblate spheroid indicating it moves predominantly perpendicular to its square-planar environment. In contrast, the comparatively small and perfectly spherical thermal ellipsoid of Zn indicates it moves very little. The ellipsoid for N is isotropic perpendicular to the bond compared to that of C, which has an oblate shape more similar to that of Ni.*

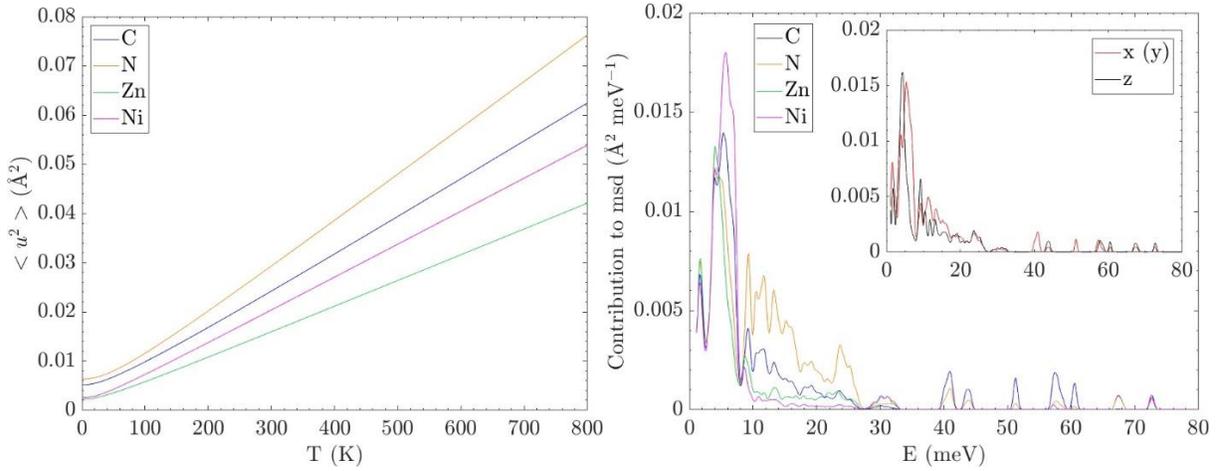

FIG. 9. *MSD with temperature for each atom type in ZnNi(CN)$_4$ (left). This represents the average of a single atom in the unit cell. The integrand of equation 3 representing the contribution to the MSD at 300 K of phonon modes with energy, E, for each atom type (center) and averaged over all atoms projected along the x (y) and z axes (right). Low-energy modes contribute the most to the MSDs. N has the largest amplitude of motion as a function of temperature arising from modes between 10 and 20 meV, as a result of greater freedom of movement around the Zn. Ni has a larger MSD than Zn, dominated by modes around 7 meV. At this energy, Ni has a larger amplitude of motion than even C and N. (see Fig. 14).*

The mode Grüneisen parameters for each phonon at $q$-points from a $40 \times 40 \times 20$ $k$-point mesh are shown in Fig. 10. The range in value over the Brillouin zone decreases rapidly with energy as the modes become more localized.

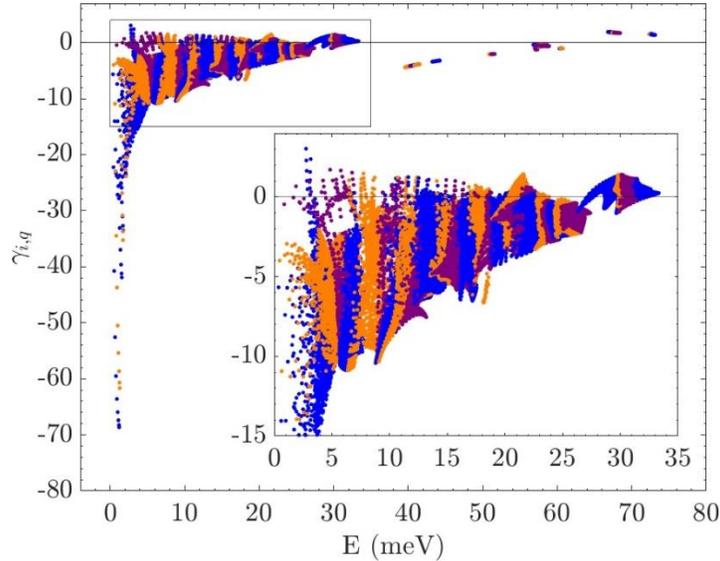

FIG. 10. *The calculated mode Grüneisen parameters, $\gamma_i$, at q-points from a $40 \times 40 \times 20$ k-point mesh over the whole Brillouin zone. A color-coded illustration for neighboring modes is used for contrast. The most negative Grüneisen parameters can be seen in the low-energy region, corresponding to the acoustic modes.*

For all the high symmetry points in the Brillouin zone, there are phonon modes with negative Grüneisen parameters of ~ −9 (Table IV). However, the lowest-energy acoustic mode at the zone boundary in the Z (0 0 ½) direction has a very pronounced mode-Grüneisen-parameter value of −79.40. The eigenvectors for this mode describe an almost rigid transverse motion of both [Zn(NC)$_4$] tetrahedra in the unit cell, resulting in a transverse motion and slight twisting of the [Ni(CN)$_4$] units. The real displacement vectors for this mode are depicted in Fig. 11.

Table IV. *The frequencies and corresponding mode Grüneisen parameters for the two modes with the most negative Grüneisen parameter along the high symmetry point in the Brillouin zone (crystallographic point group P4$_2$/mcm). The asterisk denotes acoustic modes.*

| Zone Boundary | $\omega_i$ /meV | $\gamma_i$ |
|---|---|---|
| Γ (0 0 0) | 6.91, 9.06 | −10.67, −9.62 |
| Z (0 0 ½) | 1.12*, 5.71 | −79.40, −10.57 |
| R (0 ½ ½) | 5.09*, 5.91* | −10.85, −7.71 |
| X (0 ½ 0) | 5.91*, 6.04* | −9.46, −9.28 |
| M (½ ½ 0) | 7.65*, 8.77 | −7.92, −10.45 |
| A (½ ½ ½) | 7.07*, 9.56 | −9.03, −7.54 |

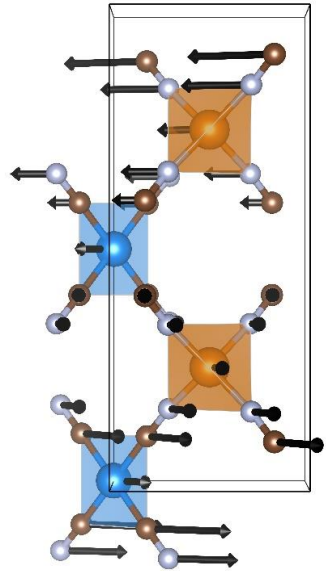

FIG. 11. *Schematic illustration of zone-boundary acoustic phonon mode with energy 1.12 meV, along Z (0, 0, ½), with the highest mode Grüneisen parameter of −79.40. The vectors represent the real displacement, however their amplitude has been exaggerated.*

The calculated dispersion curves for the ordered ZnNi(CN)$_4$ model are presented in Fig. 12 in two ways. The left-hand figure corresponds to the full dispersion relations. In the right-hand figure, a filter is applied according to the mode Grüneisen parameters highlighting which modes contribute the most to NTE.

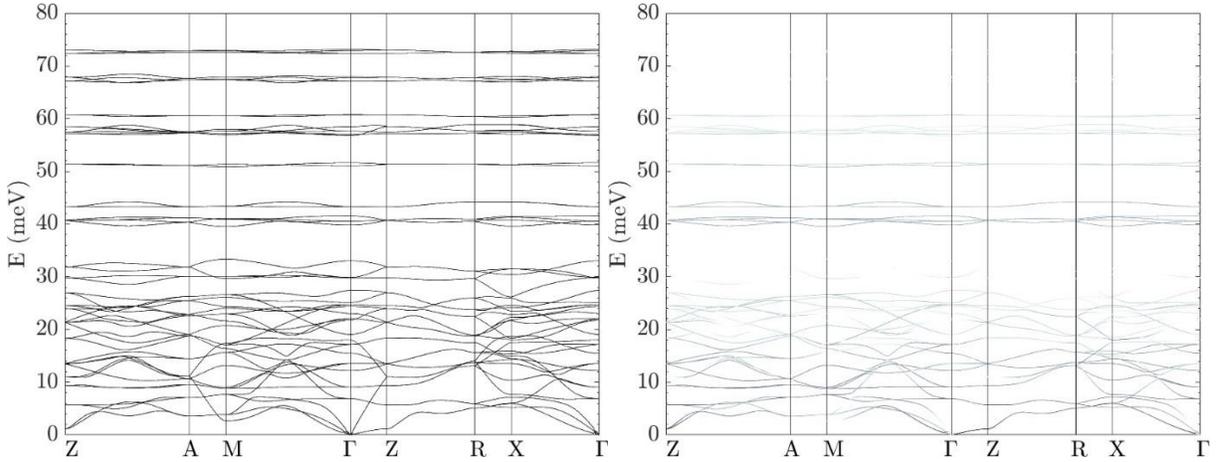

FIG. 12. *Left: The calculated phonon dispersion curves for ZnNi(CN)$_4$ along the indicated high-symmetry points (the CN stretching modes at ~280 meV are not shown). Right: The dispersion curves with grayscale color coding according to values of the mode Grüneisen parameters. Black indicates the most negative value, whereas white represents a value ≥ 0. The more visible the dispersion curve the more it contributes to NTE in ZnNi(CN)$_4$.*

Figures 7, 10 and 12, all show that the bulk of the NTE comes from optic modes below 20 meV and around 40 meV. Analysis of mode eigenvectors reveal that the optic modes below 20 meV involve rotations and deformations where the angle between the displacement of the C and N atom is less than 90°, whereas the modes between 40 and 60 meV involve rotations and deformations where this angle is greater than 90°. Zone-center modes that involve a perfect rotation of a unit are shown in Fig. 13. A view along the *c*-axis clearly shows how the transverse or rotational motion of the CN ligand results in a decrease in the *a* lattice parameter. The modes at 12.20 and 39.81 meV represent perfect RUMs. In the former (12.20 meV), neighboring tetrahedral and square-planar units rotate towards one another, and in the latter (38.81 meV), they rotate in phase away from one another. The other four modes shown represent a RUM for one unit inducing a deformation in neighboring units. Away from the zone center, the motions described occur with different relative phases.

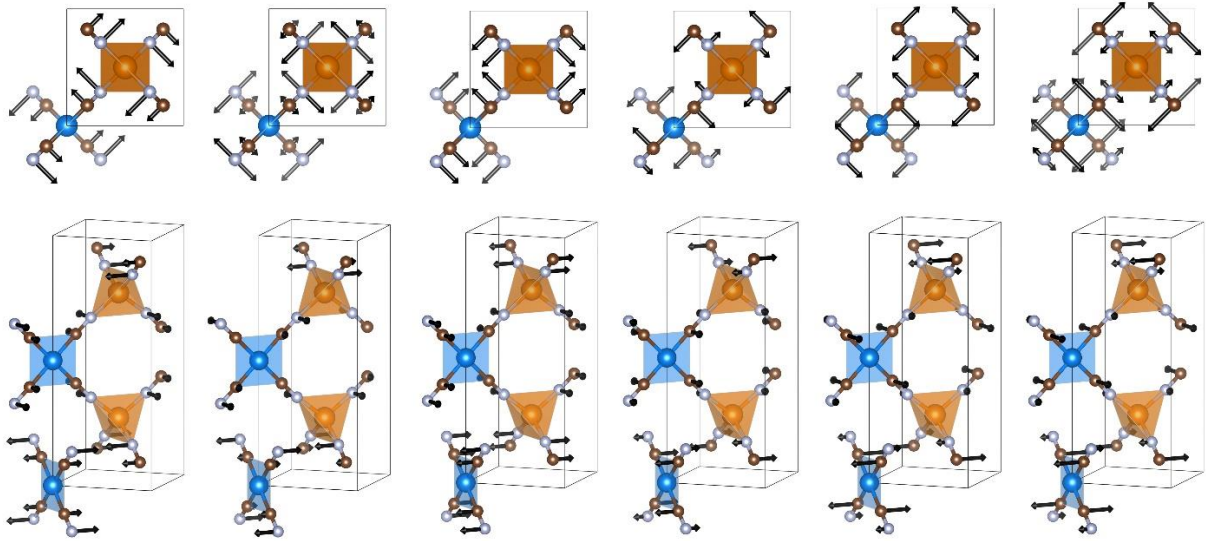

FIG. 13. *Schematic illustrations of zone center phonon modes viewed along the c-axis (top) and from the side (bottom) demonstrating how their motions shrink the ab plane. The arrows represent the real displacement vectors of the atoms. The first and fourth modes are perfect rotational RUMs. The second and sixth are RUMs for the tetrahedral units and distortions of the square planar units, whereas the third and fifth are RUMs for the square-planar units and distortions of the tetrahedra. Frequencies for each mode from left to right are (in meV): 12.20, 15.48, 19.05, 38.81, 41.55, 51.04. The corresponding mode Grüneisen parameters are: –5.029, –5.163, –0.781, –4.367, –3.880, –2.066, respectively.*

Although the rotational RUM modes described do make an important contribution to the NTE, they are not the optic modes that contribute the most. The lowest-energy optic mode, shown at the zone center in Fig. 14, involves an undulation of the framework that is not present in $Zn(CN)_2$. Due to the 2D nature of the square-planar units, they can act as flexible connectors between two Zn metals, almost like a giant ligand. Hence, their transverse motion will also bring the neighboring Zn ions closer together. This motion is neither a RUM for the [$NiC_4$] or [$ZnN_4$] units, nor for the larger tetrahedral [$Zn(NCNi)_4$] unit. This is an example of the *tension effect*, but not of a RUM.

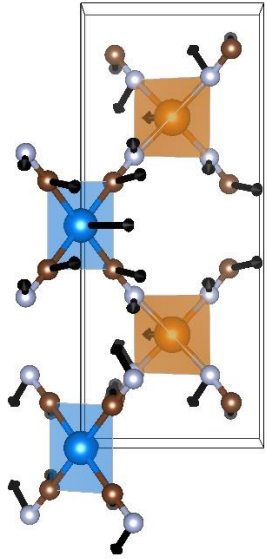

FIG. 14. *The lowest energy optic mode at the zone center, which has a frequency of 6.9 meV and Grüneisen parameter of −10.67. This mode is the optic mode that contributes the most to the NTE, but it is not a RUM. The motion is possible due to the square-planar coordination of the Ni making it only constrained in two dimensions. (The lower Ni atom is moving into the plane page.)*

# V. CONCLUSION

A polycrystalline sample of ZnNi(CN)$_4$ was synthesized using a precipitation method. Powder x-ray diffraction, density measurements, IR and Raman spectroscopy, and the experimentally determined atomic correlation function showed that the double interpenetrating framework structure was formed. The x-ray diffractogram contains broad peaks indicating poor crystallinity. Total energy calculations using a tetragonal model derived from published single-crystal study [14] confirm that the Zn atoms bond exclusively to the carbon end of the cyanide ligand. Bond lengths were extracted from the total atomic correlation function obtained using neutron scattering. The values obtained show those previously reported [14] are in error. Relaxation of the structure by *ab initio* methods yielded bond lengths matching those from the neutron diffraction data.

Inelastic neutron scattering was used to study the dynamics of ZnNi(CN)$_4$. *Ab initio* calculations under the QHA using a tetragonal model reproduced the measured phonon spectra well, validating its use for dynamical predictions of the crystalline structure. DFT calculation at 0 K of the bulk modulus yields a similar value to that of Zn(CN)$_2$, however its pressure derivative is negative. This possibly points towards a pronounced flexibility of the ordered structure, which could contribute to the broad nature of the x-ray diffraction peaks. Calculated thermal expansion coefficients predict 2D NTE with overall volume NTE in crystalline ZnNi(CN)$_4$. Analysis of the mode eigenvectors and Grüneisen parameters show that although low-energy phonons with RUM-like character do contribute to the NTE, twisted transverse motions, which are not present in Zn(CN)$_2$, play a more significant role. The motions are the result of the lower constraint on the Ni atom due to its 2D coordination geometry within a 3D framework.

The results in this work provide further evidence that framework structures containing corner-sharing geometries are likely to contain low-energy phonon modes capable of inducing NTE. When designing a new compound capable of exhibiting anomalous thermal properties, low structural constraints are effective, however they may have unforeseen effects on the crystallinity of the material. This study could be pursued further by extending the INS probe to cover a wider temperature range as well as to attempt also performing pressure-dependent measurements to gain deeper and explicit insights into both temperature and volume effects on phonon dynamics and related properties of $ZnNi(CN)_4$.

## ACKNOWLEDGMENTS

The ILL is thanked for providing beam time on the IN6 spectrometer and ISIS for providing beamtime on the GEM diffractometer. The use of the Chemical Analysis Facility (CAF) at Reading is acknowledged. S. d'A. thanks the ILL and the University of Reading for financial support.